\documentclass[twocolumn,prp,aps]{revtex4-1}
\usepackage{amsmath}
\pdfoutput=1
\usepackage{amssymb}
\usepackage{graphicx}
\usepackage{dcolumn}
\usepackage{graphicx}
\usepackage{dcolumn}
\usepackage{bm}

\begin{document}

%\preprint{Phys.Rev.Lett.}

\title{Microwave-induced magnetooscillations and signatures of zero-resistance
states in phonon-drag voltage in two-dimensional electron systems}

\author{A. D. Levin,$^1$ Z. S. Momtaz,$^1$ G. M. Gusev,$^1$, O. E. Raichev,$^2$ and A. K. Bakarov,$^{3,4}$}

\affiliation{$^1$Instituto de F\'{\i}sica da Universidade de S\~ao
Paulo, 135960-170, S\~ao Paulo, SP, Brazil}

\affiliation{$^2$Institute of Semiconductor Physics, NAS of
Ukraine, Prospekt Nauki 41, 03028 Kyiv, Ukraine}

\affiliation{$^3$Institute of Semiconductor Physics, Novosibirsk
630090, Russia}

\affiliation{$^4$Novosibirsk State University, Novosibirsk 630090,
Russia}

\date{\today}

\begin{abstract}

We observe the phonon-drag voltage oscillations correlating with the resistance oscillations
under microwave irradiation in a two-dimensional electron gas in perpendicular magnetic field.
This phenomenon is explained by the influence of dissipative resistivity modified by microwaves
on the phonon-drag voltage perpendicular to the phonon flux. When the lowest-order resistance
minima evolve into zero-resistance states, the phonon-drag voltage demonstrates sharp features
suggesting that current domains associated with these states can exist in the absence of
external dc driving.

\pacs{73.43.Qt, 73.50.Lw, 73.50.Pz, 73.63.Hs}

\end{abstract}

%\nofiles

\maketitle

The microwave-induced resistance oscillations (MIRO) [1] and the related phenomenon
of zero-resistance states [2] are observed in high-mobility two-dimensional (2D) electron
systems placed in a relatively weak magnetic field $B$ and subjected to microwave (MW)
irradiation. The underlying physics of MIRO turned out to be unexpectedly rich, and its
basic concepts have been applied to a variety of magnetotransport phenomena discovered recently [3].
The mechanisms of MIRO [3-9] originate from the scattering-assisted electron transitions
between Landau levels due to absorption and emission of the MW radiation quanta by 2D
electrons. Such processes strongly modify the energy distribution of electrons even in the
case of overlapping Landau levels and lead to an oscillating contribution to electrical
resistivity determined by the ratio of the radiation frequency $\omega$ to the cyclotron
frequency $\omega_{c}$. This physical picture is widely accepted [3], although
alternative explanations of MIRO exist [10,11]. The theory successfully describes
the period and phase of the magnetooscillations as well as the dependence of their
amplitude on temperature and MW power.

The increase in MW power enhances the amplitude of MIRO and transforms the lower-order
minima of magnetoresistance into the zero-resistance states (ZRS), the intervals of $B$
where the dissipative resistance stays equal to zero. The theories developed for
spatially homogeneous transport predict a negative resistivity in these intervals.
Since the negative resistivity means that the homogeneous current distribution
becomes unstable, it was suggested [12] that a pattern of current domains is formed in
the sample. Indeed, in this regime a change in the current is accommodated by a shift
of the domain walls without any voltage drop, so the resistance becomes zero. The
experiments [13-16] support this concept.

In spite of attention to the MIRO and ZRS phenomena, experimental studies of the
magnetotransport properties of 2D electrons under MW irradiation are almost entirely
based on the measurements of electrical resistance or conductance under dc driving.
An exception is the observation of MW-induced photovoltaic oscillations [17-19]. These
oscillations occur in the samples with built-in spatial variation of electron density
because the MW irradiation strongly modifies the conductivity while leaving the
diffusion coefficient unaffected [19]. Here we suggest that when temperature $T$
varies across the sample, the MW irradiation creates conditions which allow one to
observe, without any external dc driving, an oscillating thermoinduced voltage
proportional to the resistance and closely resembling a MIRO signal.

To illustrate the basic idea, let us represent the current density as a sum of the
drift current and the thermoelectric one [20], ${\bf j}=\hat{\sigma}{\bf E}-
\hat{\beta} \nabla T$, where $\hat{\sigma}$ and $\hat{\beta}$ are the conductivity
and thermoelectric tensors. In thermopower measurements, when ${\bf j}=0$, the
electric field ${\bf E}=\hat{\alpha} \nabla T$ is determined by the thermopower tensor
$\hat{\alpha}=\hat{\rho} {\hat \beta}$, where $\hat{\rho}=(\hat{\sigma})^{-1}$ is the
resistivity tensor. In the regime of classically strong magnetic fields relevant for
high-mobility electrons, the longitudinal thermopower is $\alpha_{xx}=\rho_{xy}
\beta_{yx}+\rho_{xx}\beta_{xx}\simeq \rho_{xy}\beta_{yx}$, since the second term is
small as $(\omega_c\tau_{tr})^{-2}$, where $\tau_{tr}$ is the transport time. In the
transverse thermopower, $\alpha_{xy}=\rho_{xy}\beta_{yy}+\rho_{xx}\beta_{xy}$, both
terms are important. The main
contribution to $\hat{\beta}$ in 2D systems comes from phonon drag mechanism [21-23],
when electrons are driven by a frictional force between them and the phonons
propagating along the temperature gradient. In these conditions, two terms in
$\alpha_{xy}$ compensate each other [20] and $\alpha_{xy}=0$ in the classical
regime. Under MW irradiation the dissipative resistivity $\rho_{xx}$ is strongly modified
while the Hall one, $\rho_{xy}$, remains unchanged. The coefficients $\beta_{yy}$
and $\beta_{xy}$ are not modified by microwaves as strongly as $\rho_{xx}$ [24]. Thus,
the terms in $\alpha_{xy}$ no longer compensate each other, and the voltage developing
in the direction perpendicular to the phonon flux should have an oscillating component
proportional to the MW-induced part of dissipative resistivity.

In this Letter, we report experimental studies of phonon-drag voltage (PDV) in
high-mobility 2D electron gas in four-terminal devices. We measure the dissipative
resistance $R_{xx}$ simultaneously with PDV and observe correlations of the PDV
oscillations with MIRO under MW excitation. More important, we see sharp features
in PDV corresponding to ZRS in $R_{xx}$, which apparently indicate that the
ZRS domains can exist in the absence of external dc driving. A theoretical
consideration confirms the direct relation between PDV and resistance.

We have studied narrow (14 nm) quantum wells with electron density $n_s\simeq
10^{12}$ cm$^{-2}$ and a mobility $\mu=2\times10^6$ cm$^2$/V s,
at $T=1.5$ K. The 2D electron gas occupies a circular central part (diameter 1 mm)
and four long (length $\simeq 5$ mm, width 0.1 mm) arms ending with the voltage probes.
The measurements have been carried out in a VTI cryostat with a waveguide to deliver
MW irradiation (frequency range 110 to 170 GHz) down to the sample. The heater
placed symmetrically between the arms 1 and 2 at a distance of 4.1 mm from the
center (see Fig. 1) generates phonon flux. The voltages induced by this flux
were measured by a lock-in method at the frequency of $2f_0=54$ Hz both in the
longitudinal, $V_{14}$ and $V_{23}$, and in the transverse, $V_{12}$ (hot side)
and $V_{43}$ (cold side) configurations. Without powering the heater no photovoltage
was observed. The thermovoltage increases linearly with heater power. We
find the electron temperature near the heater and heat sink by the
2-probe measurements (contacts 1-2 and 3-4), exploiting the amplitude of
the Shubnikov-de Haas (SdH) oscillation. The difference in the electron temperature
between hot and cold sides is found $\Delta T\simeq0.3$ K at the lattice
temperature $T=1.5$ K. Several devices from the same wafer have been studied. The
magnetoresistance (Fig. 1) was measured as a response $V_{14}$ to the current
injected through the contacts 2 and 3. The ZRS is seen at $T$ below 4.2 K.
Similar results are obtained for the other contacts.

\begin{figure}[ht]
\includegraphics[width=9.cm]{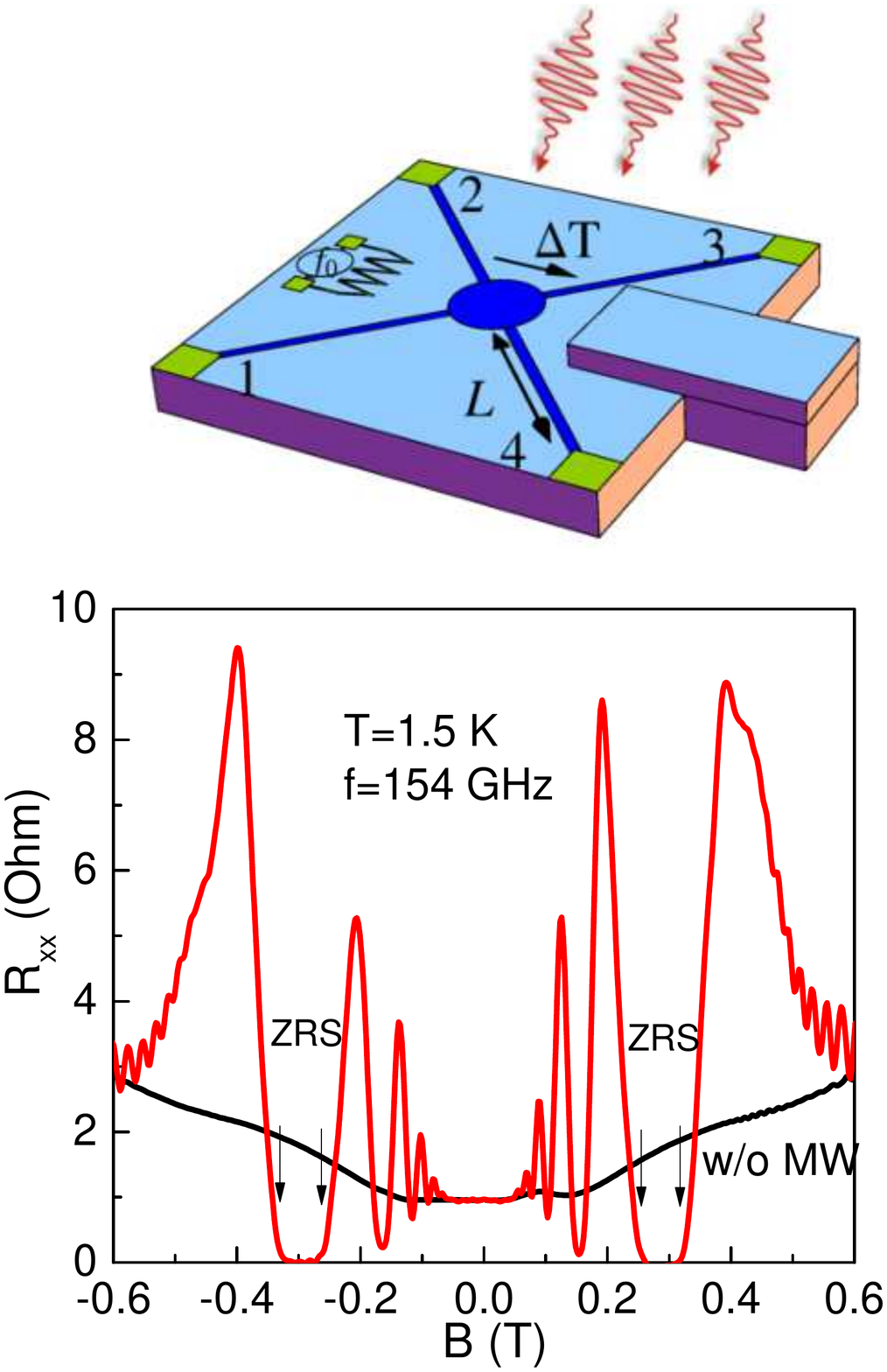}
\caption{
(Color online) Sample geometry and longitudinal resistance without and with MW
irradiation (154 GHz) as a function of magnetic field. Arrows show ZRS region.}
\end{figure}

The magnetooscillations of $V_{23}$ and $V_{12}$, both with MW irradiation
and without it (dark signal), are shown in Figs. 2 and 3. The transverse cold-side
voltage $V_{43}$ is much weaker than $V_{12}$, though shows the same periodicity
as $V_{12}$. Both dark voltages $V_{23}$ and $V_{12}$ demonstrate acoustic
magnetophonon oscillations whose period is determined by the ratio $2 k_F
s_{\lambda}/\omega_c$, where $k_F=\sqrt{2\pi n_s}$ is the Fermi wavenumber
and $s_{\lambda}$ is the sound velocity for phonon mode $\lambda$. These
oscillations due to resonant phonon-assisted backscattering of electrons
were observed previously [23]. The MW irradiation enhances both $V_{12}$ and
$V_{23}$ by adding oscillating contributions odd in $B$. The positions
of the peaks and minima of these PDV oscillations coincide with those of MIRO.
The MW-induced contributions to $V_{23}$ and $V_{14}$ have opposite signs.

According to the general symmetry properties of thermoelectric coefficients, the
odd in $B$ voltages develop in the direction perpendicular to the temperature
gradient or phonon flux. Thus, the odd in $B$ behavior of $V_{12}$ is expectable,
while the appearance of odd in $B$ contributions to $V_{23}$ and $V_{14}$ may look
surprising. Explanation of this fact, however, is straightforward. The position of
the heater between the long radial arms of the device (see Fig. 1)
makes it clear that there is no homogeneous unidirectional phonon flux in the 2D area
of the device. The phonons coming from the heater cross the arms attached to probes 1
and 2 in the directions perpendicular to these arms, so the voltages $V_{23}$ and
$V_{14}$, apart from the longitudinal (even in $B$) phonon-drag contributions, contain
significant transverse (odd in $B$) contributions. Since the phonons come to the
arms 1 and 2 from different sides, the transverse contributions to $V_{23}$ and $V_{14}$
should have different signs, in agreement with our observation. Therefore, we identify
the observed MW-induced voltages as a result of transverse phonon-drag effect (spatial
redistribution of electrons in the direction perpendicular to the phonon flux) which
is strongly enhanced because of the influence of microwaves on the resistance, as
explained in the introduction.

\begin{figure}[ht]
\includegraphics[width=9.cm]{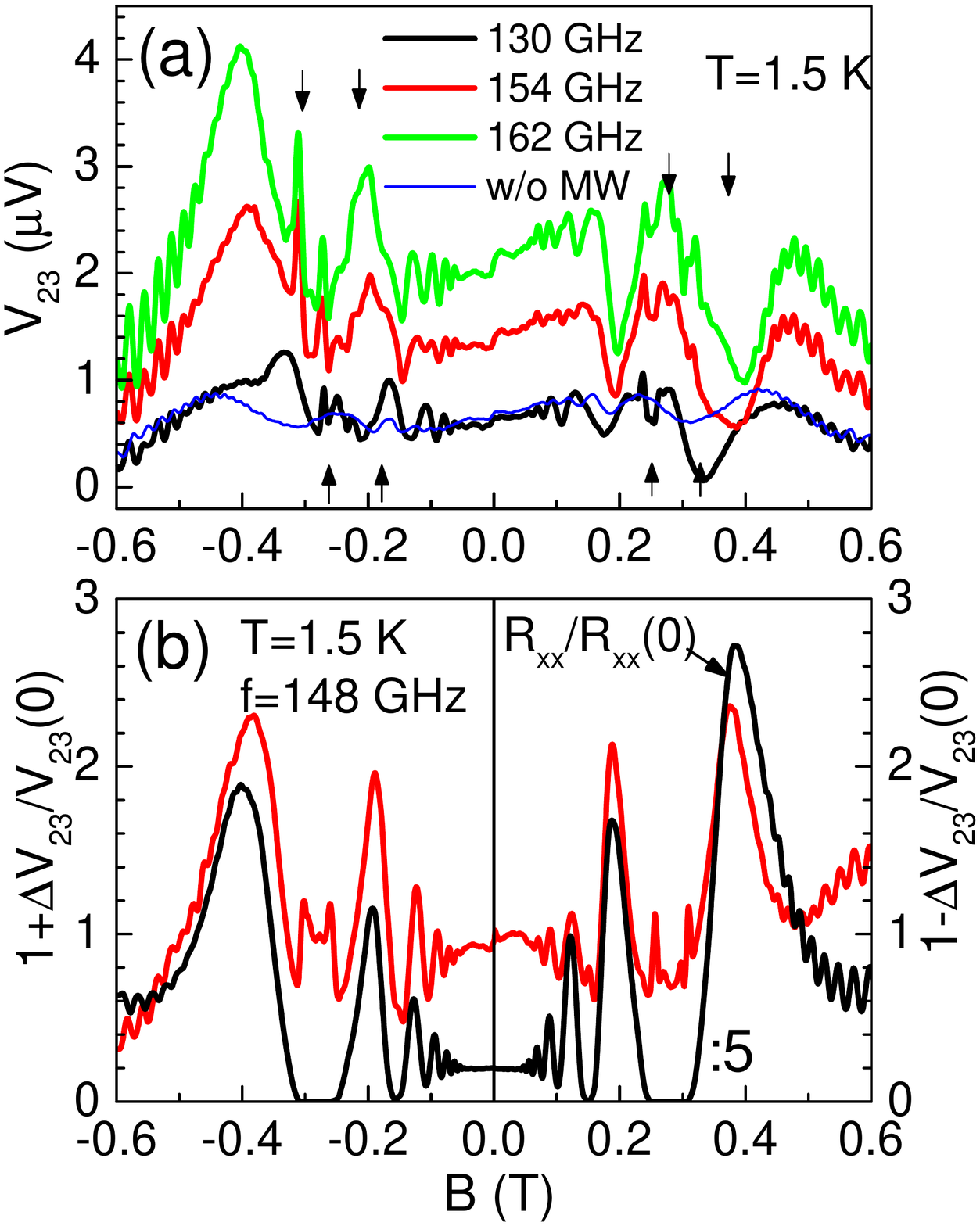}
\caption{
(Color online) (a) Magnetic-field dependence of the longitudinal PDV $V_{23}$ without
and under MW irradiation for different microwave frequencies (shifted up for higher
frequencies). Arrows show ZRS region. (b) PDV oscillations vs MIRO at 148 GHz. For
clarity of the comparison, the sign of $\Delta V_{23}\equiv V_{23}(B)-V_{23}(0)$ is
inverted at $B>0$ and the resistance is scaled down by the factor of 5.}
\end{figure}
\begin{figure}[ht]
\includegraphics[width=9.cm]{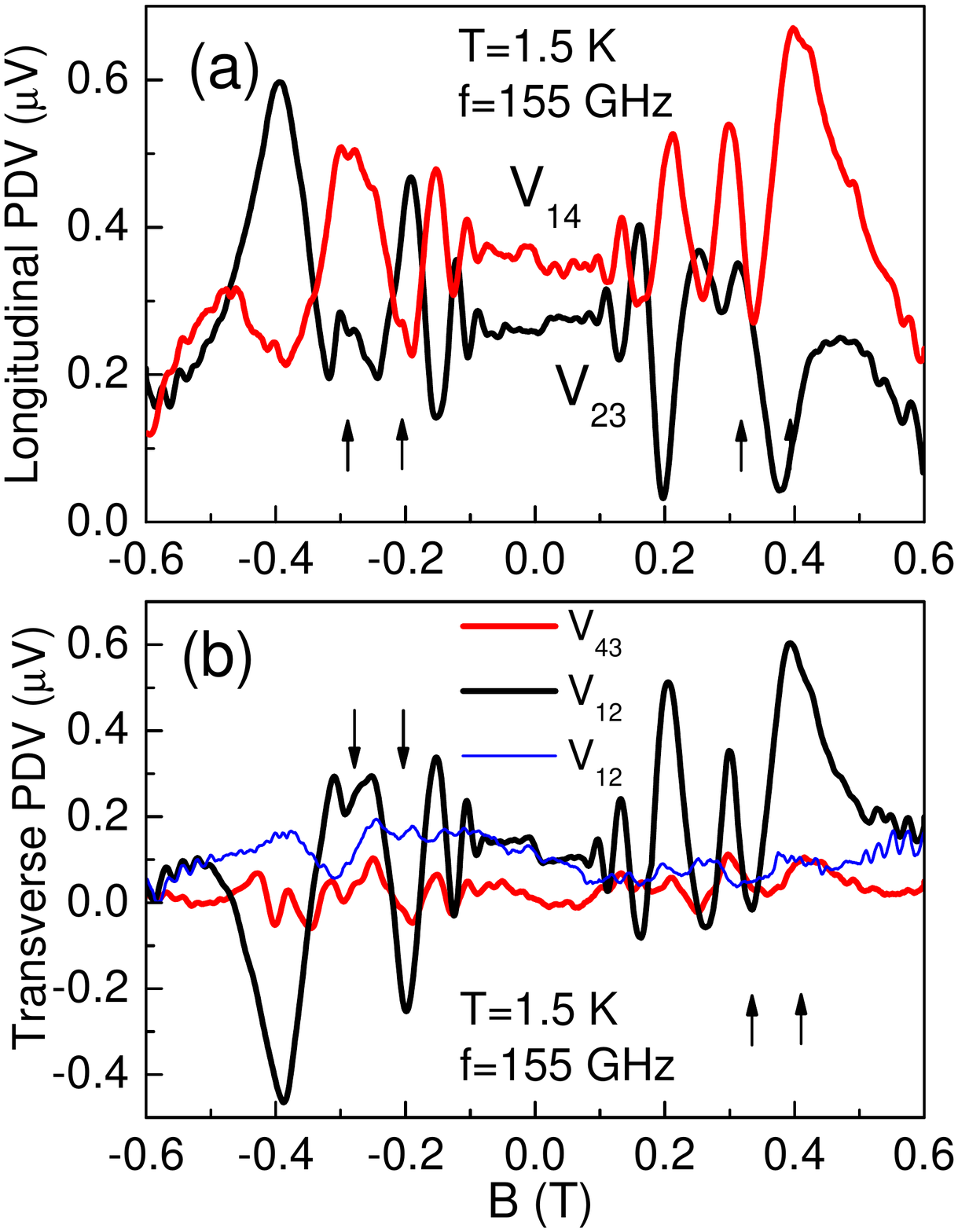}
\caption{
(Color online) (a) Magnetic-field dependence of $V_{23}$ and $V_{14}$
under MW irradiation. The MW-induced contributions to these voltages
have opposite signs. (b) Transverse PDV $V_{12}$ (high-amplitude)
and $V_{43}$ (low-amplitude) under MW irradiation. Thin line: $V_{12}$
without MW irradiation.}
\end{figure}

To get more insight into the physics of the observed phenomenon, it is important to
estimate the PDV theoretically. We believe that the drag in our experiment is mostly caused
by high-energy ballistic phonons emitted from the heater in different directions [25],
based on the following observations. First, the voltage $V_{43}$ measured far away
from the heater is much smaller than the voltage $V_{12}$, so the proximity
to the heater is essential for the PDV signal. Second, the amplitude of magnetophonon
oscillations of PDV in the absence of MW irradiation is much larger than expected
for temperature $T=1.5$ K in view of exponential suppression of backscattering
in the Bloch-Gruneisen regime, $k_B T \ll 2\hbar k_F s_{\lambda}$, so the effective phonon
temperature $T_{ph}$ must be considerably higher than $T$. The ballistic phonon model is
reliable because of large (1 mm) mean free path lengths of phonons in GaAs at low $T$ [26].
Using the basic formalism for calculation of thermoelectric response in quantizing
magnetic field [24] in application to this model [25], one may represent the PDV
between the contacts $i$ and $j$ in the form
\begin{equation}
V_{ij}= \gamma^L_{ij} E_L + \gamma^T_{ij} E_T,
%1
\end{equation}
where $E_L$ and $E_T$ are the electric fields developing as phonon-drag responses to a
homogeneous unidirectional phonon flux in the directions along and perpendicular to
this flux, respectively, while the lengths $\gamma^L_{ij}$ and $\gamma^T_{ij}$ depend on
the sample geometry. The PDV are formed by mixing of the longitudinal
(even in $B$) and transverse (odd in $B$) phonon-drag contributions described
by the following expressions obtained in the regime of weak Landau quantization [25]:
\begin{equation}
E_L=F + 2d^2 G,~~E_T = \left( 2d^2 G - F \delta \rho_{xx}/\rho_{0} \right)/\omega_c \tau_{tr},
%2
\end{equation}
where $\delta \rho_{xx}=\rho_{xx}-\rho^{(0)}_{xx}$, $\rho_{0}=m/e^2n_s\tau_{tr}$,
$\rho^{(0)}_{xx}=\rho_{0}(1+2d^2)$ is the resistivity in the absence of MW irradiation,
$d=\exp(-\pi/|\omega_c|\tau)$ the Dingle factor, $\tau$ the quantum lifetime of
electrons and $\tau_{tr}$ the transport time. The quantity $F$ does not depend on
$B$ while $G$ is a function of $B$ describing the magnetophonon oscillations of PDV [25].
In the absence of MW irradiation, $E_L \gg E_T$ in the relevant regime of classically strong
magnetic fields, $|\omega_c| \tau_{tr} \gg 1$, so the dark PDV $V_{23}$ is governed by $E_L$
and is even in $B$. Under MW irradiation, $E_T$ increases dramatically because of the large
ratio $\delta \rho_{xx}/\rho_{0}$ and gives large, odd in $B$ contributions to all
measured PDV.

\begin{figure}[ht]
\includegraphics[width=9.cm]{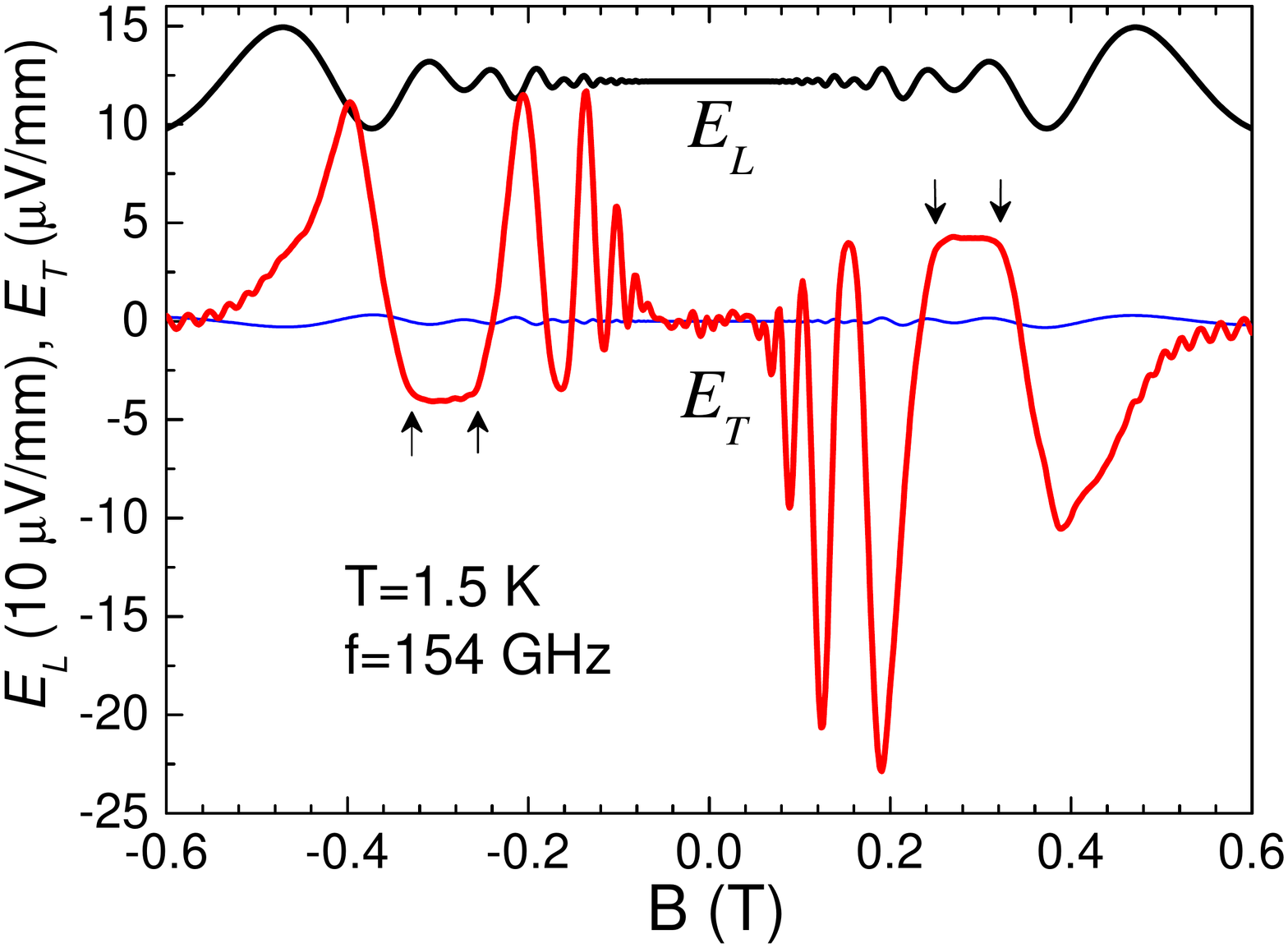}
\caption{
(Color online) Calculated magnetic-field dependence of the fields $E_L$ and $E_T$, the
latter is plotted both with MW irradiation and without it (thin line). The function
$\delta \rho_{xx}/\rho_0$ entering $E_T$ is extracted from the experiment.}
\end{figure}

The theoretical plots of $E_L$ and $E_T$ for our sample are shown in Fig. 4.
The effective temperature of ballistic phonons, $T_{ph} \simeq 4$ K, is estimated
from the amplitude of magnetophonon oscillations in $V_{23}$. The magnetic-field
dependence of $E_T$ shows a strong oscillating enhancement under MW irradiation. To plot it,
we substituted the experimental dependence of $\delta \rho_{xx}/\rho_0$ into Eq. (2).
Similar results are obtained using theoretical dependence of $\delta \rho_{xx}/\rho_0$ [7].
Our estimates of $V_{23}$ and $V_{12}$ based on the calculated $E_L$ and $E_T$ are in the
general agreement with experiment (note that in Eq. (1) one should take into account that
both $\gamma^L_{ij}$ and $\gamma^T_{ij}$ are small compared to the device size [25]).
Therefore, the theory confirmes that the observed MW-induced changes of the PDV are caused
by the effect of microwaves on the dissipative resistance.

In the ZRS regions, the experimental PDV shows a complex and diverse behavior that cannot
be explained within the theory given above. Our observations reveal abrupt changes
of the drag voltages in obvious correlation with the ZRS in $R_{xx}$, see Fig. 2 (b). Most
often, the PDV, as a function of $B$, jumps at the beginning and at the end of the ZRS
region, and more sharp features also appear within this region. We attribute
this behavior to a transition from the homogeneous transport picture to the domain structure
specific for the ZRS, since such a transition is accompanied with switching between different
distributions of the electric field in the 2D plane [12,15]. We emphasize that in our
experiment this transition occurs in the unusual conditions, when external dc driving
is absent. Nevertheless, this fact rests within the general theoretical picture of ZRS [12],
because the instability of the spatially homogeneous state is irrelevant to the presence of
dc driving and requires only the negative conductivity created, for example, by MW
irradiation. The resulting domains may carry electric currents, and the domain arrangement
should provide zero currents through the contacts. The details of such domain structures
are not clear and require further studies.

In summary, we observe MW-induced magnetooscillations of the phonon-drag voltage in GaAs
quantum wells, correlating with the behavior of electrical resistance. The effect is
described in terms of the sensitivity of transverse drag voltage to the dissipative
resistivity modified by microwaves. The behavior of phonon-drag voltage in the zero
resistance regime can be viewed as a signature of current domain state.
Such MW-induced thermoelectric phenomena may show up in other 2D systems. The
magnetothermoelectric measurements are therefore established as a tool to study the
influence of MW radiation on the properties of 2D electrons and to gain complementary
information about the MIRO and ZRS regime.

The financial support of this work by FAPESP, CNPq (Brazilian agencies) is acknowledged.

%\end{references}

\end{document}